\documentclass{desyproc}
\usepackage{psfrag}
\usepackage{wrapfig}

\newcommand{\OP}{\omega_\mathrm{_P}}

\renewcommand\({\left(}
\renewcommand\){\right)}
\renewcommand\[{\left[}
\renewcommand\]{\right]}

\newcommand{\be}{\begin{equation}}
\newcommand{\ee}{\end{equation}}
\def\bea{\begin{eqnarray}}
\def\eea{\end{eqnarray}}

\begin{document}
\title{A Hidden Microwave Background ?\\
\large\centering
- signatures of photon-WISP oscillations in the CMB -}

\author{{\slshape Javier Redondo}\\[1ex]
Max-Planck-Institut, F\"ohringer Ring 6, D-80805 M\"unchen, Germany\\
}
\contribID{redondo\_javier\_2}

\desyproc{DESY-PROC-2009-05}
\acronym{Patras 2009} 
\doi  

\maketitle

\begin{abstract}
The existence and cosmological signatures of a relic background of very weakly interacting sub-eV particles (WISPs), 
produced by photon-WISP oscillations is reviewed.
\end{abstract}

\section{Introduction}

Very weakly interacting sub-eV particles (WISPs) appearing in a hidden sector of nature, i.e. a sector of particles carrying no standard model charges, can mix with photons.
This is the case of the standard graviton, and also of hypothetical particles such as axions, axion-like-particles (ALPs)~\cite{Raffelt:1987im} or hidden photons ($\gamma'$)~\cite{Okun:1982xi}.
In the last case, the mixing can be provided by a non-diagonal kinetic term, a so-called kinetic mixing that after a field redefinition appears as $\gamma-\gamma'$ mass mixing~\cite{Ahlers:2007rd}. 
In all the previous cases, mixing cannot occur at tree level (these WISPs have spin different from 1). 
However, the existence of WISP couplings to \emph{two} photons can produce an effective mixing term in an background magnetic field.

The WISP-photon mixing gives a non-diagonal contribution to the mass matrix which no longer allows photons  to be propagation eigenstates. This leads to the phenomenon of photon oscillations and photon disappearance, analogously to the neutrino case.
The $\gamma\leftrightarrow$WISP conversion probability as a function of propagation length $L$  in a medium of index of refraction $n$ is given by
\be
\label{prob}
P(\gamma\to \phi)=
\frac{4\delta^2}{(m_\phi^2-m_\gamma^2)^2+4\delta^2}
\sin^2 \(\frac{((m_\phi^2-m_\gamma^2)^2+4\delta^2)^{1/2}L}{4\omega}\)=\sin^2 2\theta \sin^2 \(\frac{\pi L}{L_{\rm osc}} \)
\ee
where $m_\phi$ is the WISP mass and $m_\gamma^2 \simeq -2 \omega^2 (n-1)$ is an effective photon mass with $\omega$ the photon frequency.

The mixing term $\delta$ depends on the particular WISP.
For gravitons we have $\delta = \sqrt{32\pi} B \omega/M_{\rm Pl}$  where $B$ is the component of the external magnetic field perpendicular to the photon propagation direction and the Planck mass is $M_{\rm Pl}=1.22\times 10^{19}$ GeV.
Axions and ALPs have $\delta = g B \omega$, with $g$ the two photon coupling (widely discussed in this workshop) and \emph{they only mix with one photon polarization}.
For hidden photons we have $\delta_{\gamma'}=\chi m_{\gamma'}^2$ with $\chi$ the kinetic mixing parameter. 
String motivated models give plausible values of $\chi$ in the $10^{-16}\sim 10^{-3}$ range~\cite{Dienes:1996zr
}.

Note that in vacuum, $\gamma\leftrightarrow$WISP oscillations can be very suppressed if the WISP mass is much larger than $\delta$. 
However, inspection of Eq.~(\ref{prob}) reveals that, even in this case, the amplitude of oscillations can be made maximal (1, indeed) in a medium which gives a photon effective mass that matches the WISP mass ($m_\gamma=m_\phi$) producing a resonant effect.

A typical experiment looking for $\gamma\leftrightarrow$WISP oscillations could be:
1) Take a very intense and well understood light source, 2) make it propagate the longest possible distance through 3) a medium whose index of refraction is fairly homogeneous and tunable and 4) try to detect a distortion in the light after propagation. 
5) If no distortion is observed, tune another index of refraction (which makes resonant another WISP mass) and try again.

The cosmic microwave background provides an excellent source for such an experiment.
First, it is a very well measured and understood source: the black-body nature of its spectrum is well accounted in terms of QED and standard thermodynamics in the $\Lambda$CDM cosmological model, and was measured to a precision of $10^{-4}$ by the FIRAS on the COBE satellite~\cite{Fixsen:1996nj}. More precise measurements are also under consideration~\cite{Fixsen:2002,Kogut:1996zb}.
Second, the CMB photons travelled through the very homogeneous primordial plasma. 
For most of the time, its index of refraction was smaller than one, which is a paramount requirement since otherwise $m_\gamma^2$ would be negative and a resonance impossible.
Moreover, as the universe slowly expanded, the plasma became increasingly sparse and the the index of refraction decreased accordingly. 
As we will see, the photon effective mass swept all possible WISP masses. 
We can therefore look for signatures of WISPs regardless of their mass.
Finally, CMB photons travel the longest conceivable distance for an experiment, basically the size of the universe, enhancing enormously the conversion probabilities and thus the WISP signatures.

\section{Photon mass in the early universe}

In our studies we have used a simplified model, yet containing the main features, for the effective photon mass in the primordial plasma. The typically dominant part is a positive contribution from the free electrons through the plasma frequency $\omega_{\rm P}$, while a negative (frequency-dependent) part from electrons bound in H atoms plays a role in special cases
\be
m_\gamma^2 = \OP^2(X_e) \times\[1-0.0073 \(\frac{\omega}{\rm eV}\)^2 \(\frac{1-X_e}{X_e}\)\]\ ,
\ee
where $\OP(X_e) \simeq 1.6\times 10^{-14}(1+z)^{3/2}X^{1/2}_e$~eV is the average plasma frequency in the current $\Lambda$CDM model and $X_e(z)$ is the hydrogen ionization fraction as a function of redshift (taken from~\cite{Seager:1999bc} for recombination and modelled  around redshift $z\sim 7-10$ for reionization).

\section{Transition probability in an expanding universe}

The transition probability in Eq.~(\ref{prob}) is only valid in an homogeneous medium. In the expanding universe we should account for the variation in time or redshift of the different physical quantities and the problem becomes substantially more complex. Moreover, if the resonance happens before recombination photon scattering can be important during the resonance and shall be included. 
The latter case was explored in a first paper focusing in the $\gamma'$ case~\cite{Jaeckel:2008fi} while resonances after recombination were presented in~\cite{Mirizzi:2009iz}. The ALP case was developed in~\cite{Mirizzi:2009nq}. 
While every case is different, it turns out that the results are equivalent in the regime of small transition probability. In these proceedings we present another way of reaching the same result by using the perturbative solution of the dispersionless equations of motion of the $\gamma$-WISP system as presented in~\cite{Raffelt:1987im}. 
Using the equivalence between length, time and redshift infinitesimals $dL\simeq dt = H^{-1}(1+z)^{-1}d z$ ($H$ the expansion parameter) we can write 
\be
P(\gamma\to \phi) =
\left|
\int
dt\,  \frac{\delta (t)}{2\omega} {\rm Exp} \left\{\i \int^t dt' \frac{m_\phi^2-m_\gamma^2(t')}{2\omega} \right\} \right|^2\simeq
 \pi \frac{\delta^2}{m_\phi^2 \omega H} \left|\frac{d \log m_\gamma^2}{d \log (1+z)}\right|_{z=z_r}^{-1} 
\ee
where for evaluating the integral we have used a saddle point approximation so that all quantities are to be evaluated in the resonance point $m_\phi=m_\gamma(z_r)$. 

Note that due to the H refraction term several resonances occur for small masses and large frequencies~\cite{Mirizzi:2009iz,Mirizzi:2009nq}. There is however a dominant one for which the last expression makes full sense, but this depends on the specific WISP. Fixing the WISP mass and neglecting the log derivative which amounts an O(1) factor we find that: 
1) in the hidden photon case $P\propto (\omega H)^{-1}$ which always decreases with redshift so the \emph{latest} resonance is the most relevant and 
2) in the ALP or graviton case $P\propto B^2\omega/H$ so the \emph{earliest} resonance dominates\footnote{Primordial magnetic fields usually increase with redshift faster than $(1+z)$, indeed in our studies we used the most conventional assumption that $B=B_0(1+z)^2$.}.

\section{Signatures of a hidden CMB}

The early $\gamma\to$WISP conversions can leave different footprints in the CMB depending on when the resonant conversion happens. The CMB is unprotected from spectral distortions below a temperature $T\sim$ keV and the $\gamma\to$WISP conversions are frequency dependent so they generally distort the blackbody shape. A careful\footnote{Whenever the resonance happens before recombination we included the re-thermalization processes of the photon spectrum.} $\chi^2$ analysis of the FIRAS monopole results allowed us to set strong constraints on hidden photons and ALPs with masses smaller than $\sim 0.2$ meV. Beyond this mass, the $\gamma\to$WISP resonance happens when photons can regain a thermal distribution by interacting with the primordial plasma. This of course makes the FIRAS bounds disappear. 
Nevertheles, the WISPs produced contribute to the dark matter of the universe and therefore affect structure formation.  
Their oscillation origin makes these WISPs to have a similar spectrum than photons, so they are in fact hot dark matter relics. As such, they behave in a completely similar fashion to the standard neutrinos by free-streaming out of the primordial over-densities and suppressing the power spectrum at small scales. 
Their effects can be included in the number of effective neutrino species
\be
N_\nu^{\rm eff}(x)=\frac{N_\nu}{1-x}+\frac{8}{7}\frac{x}{1-x}
\left(\frac{11}{4}\right)^{4/3}
\ee
where $x=\rho_\phi/\rho_\gamma$ is the fraction of the original photon density converted into WISPs during the resonance and $N_\nu$ is the effective number of neutrinos before the resonance. Comparing the value of $N_\nu^{\rm eff}$ recently inferred from WMAP5, other anisotropy probes, large scale structure surveys and supernova data with the standard value $N_\nu = 3.046$ gives $x<0.2$ with $95\%$ C.L. 
This limit translates into severe constraints for the $\gamma\to$WISP mixing.

\begin{figure}[t]
\centering
{
\psfragscanon
\psfrag{a1}[][l][0.6]{post-recombination}
\psfrag{a2}[][l][0.6]{weak-coupling}
\psfrag{a3}[][l][0.6]{ \hspace{0.2cm}$\mu$}
\psfrag{a4}[][l][0.6]{ \hspace{0.3cm}$\Delta N_{\rm eff}$}
\psfrag{malp}[][l][0.7]{$m_{\phi}$ [eV] }
\psfrag{galp}[][l][0.7]{$g \langle B^2 \rangle^{^{1/2}}\times 10^{10}$ GeV $\times$ nG\vspace{1cm}}
\includegraphics[width=7cm]{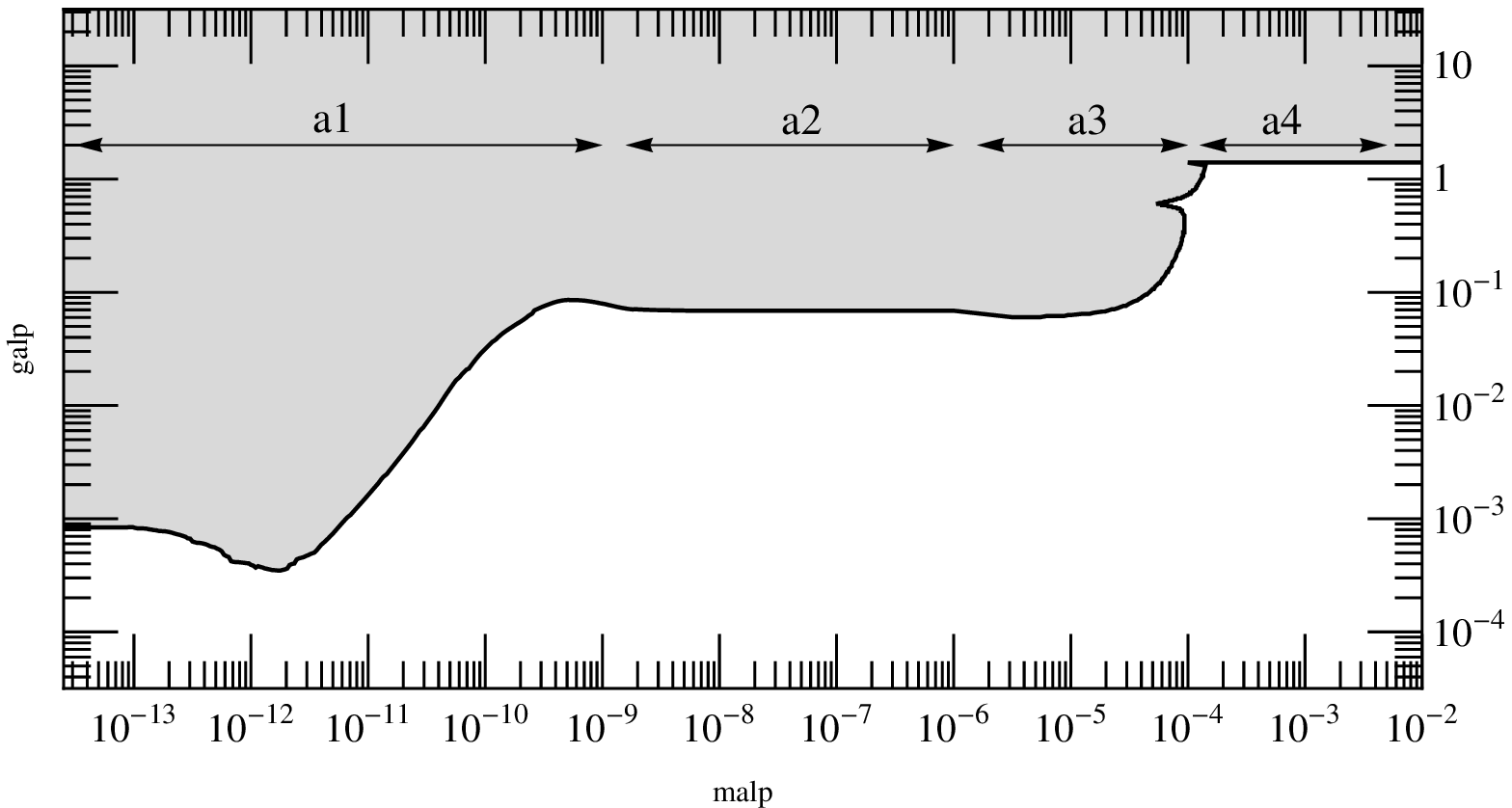} 
}
{\psfragscanon
\psfrag{a}[][l][0.7]{$m_{\gamma'}$ [eV]}
\psfrag{b}[][l][0.7]{$\chi$}
\includegraphics[width=7cm]{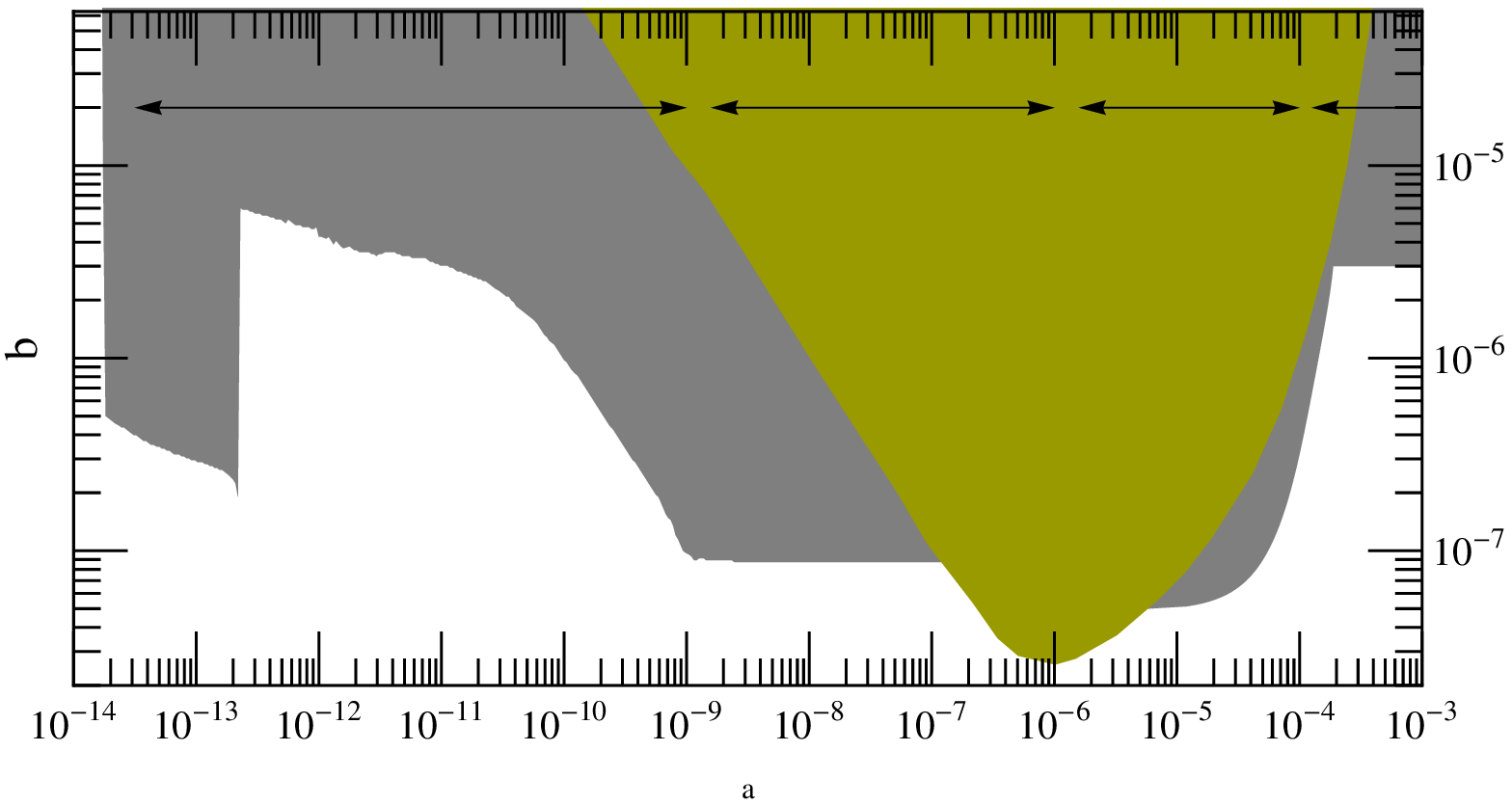}
}
\vspace{-0.35cm}
\caption{\footnotesize Constraints in the ALP (left) and hidden photon (right) parameter space from the signatures on the CMB left by a primordial $\gamma\to$WISP resonant conversion. $\Delta N_\nu^{\rm eff}$ is bounded by the WMAP5 power spectrum while for lower masses the bounds come from the FIRAS blackbody measurements.
In the hidden photon case, the yellowish region is excluded by experimental searches of deviations of the Coulomb's law~\cite{Bartlett:1988yy}.
}
\label{Fig:bounds}
\vspace{-0.35cm}
\end{figure}

A summary of the bounds obtained in~\cite{Jaeckel:2008fi,Mirizzi:2009iz,Mirizzi:2009nq} is shown in Fig.~\ref{Fig:bounds}. Note that for ALPs we only can constraint the product of the coupling times a sky averaged magnetic field during the resonance $g\langle B^2\rangle^{1/2}$. 
These results are of little use currently but can eventually turn into a fabulous diagnosis tool in the case of the discovery of an ALP, in which case one could constraint the existence of primordial magnetic fields (PMFs). Other posibility is that PMFs are discovered by other means in which case out bound will constraint the $\gamma$-ALP coupling. Bounds on $g$ lie around the $10^{-10}$ GeV$^{-1}$ ballpark while those on PMFs are slightly above the nG. If the discovery of any of those is experimentally around the corner our bounds on the other can be very relevant. 
This seems to imply that the detection of both $g$ and PMFs is very unlikely in the short term, specially for very small ALP masses.
The graviton case can be read from the ALP graph when $m_\phi\to 0$. 
Since the coupling is known one obtains a bound on the primordial field intensity of $69\ \mu$G. 
In the hidden photon case, our bounds complement and typically beat the previous bounds from modifications of the Coulomb's law.

\section*{Acknowledgements}

I would like to thank the organisers of this very interesting meeting and acknowledge support of the DFG cluster of excellence EXC 153 ``Origin and Structure of the Universe''.


\begin{footnotesize}



\begin{thebibliography}{10}

\bibitem{Raffelt:1987im}
G.~Raffelt and L.~Stodolsky,
{\em Phys. Rev.} {\bf D37} (1988)  1237.

\bibitem{Okun:1982xi}
L.~B. Okun,
{\em Sov. Phys. JETP} {\bf 56} (1982)  502.

\bibitem{Ahlers:2007rd}
M.~Ahlers, et al. \href{http://dx.doi.org/10.1103/PhysRevD.76.115005}{{\em Phys. Rev.} {\bf
  D76} (2007)  115005},
\href{http://arxiv.org/abs/0706.2836}{{\tt arXiv:0706.2836 [hep-ph]}}.

\bibitem{Dienes:1996zr}
K.~R. Dienes {\emph et al.}
{\em Nucl. Phys.} {\bf B492}  (1997)  104--118,
\href{http://arxiv.org/abs/hep-ph/9610479}{{\tt hep-ph/9610479}}; 
S.~A. Abel et al.
  \href{http://dx.doi.org/10.1016/j.nuclphysb.2004.02.037}{{\em Nucl. Phys.}
  {\bf B685} (2004)  150--170},
\href{http://arxiv.org/abs/hep-th/0311051}{{\tt arXiv:hep-th/0311051}}.
, 
  \href{http://dx.doi.org/10.1016/j.physletb.2008.03.076}{{\em Phys. Lett.}
  {\bf B666} (2008)  66--70},
\href{http://arxiv.org/abs/hep-ph/0608248}{{\tt arXiv:hep-ph/0608248}}.
and 
  \href{http://dx.doi.org/10.1088/1126-6708/2008/07/124}{{\em JHEP} {\bf 07} (2008)  124},
\href{http://arxiv.org/abs/0803.1449}{{\tt arXiv:0803.1449 [hep-ph]}};
M.~Goodsell, \emph{et al.}
  \href{http://dx.doi.org/10.1088/1126-6708/2009/11/027}{{\em JHEP} {\bf 11}
  (2009)  027},
\href{http://arxiv.org/abs/0909.0515}{{\tt arXiv:0909.0515 [hep-ph]}}.

\bibitem{Fixsen:1996nj}
D.~J. Fixsen {\em et al.}, \href{http://dx.doi.org/10.1086/178173}{{\em
  Astrophys. J.} {\bf 473} (1996)  576},
\href{http://arxiv.org/abs/astro-ph/9605054}{{\tt arXiv:astro-ph/9605054}}.

\bibitem{Fixsen:2002}
D.~J. Fixsen and J.~C. Mather, {\em ApJ} {\bf 581} (2002)  817--822.
\bibitem{Kogut:1996zb}
A.~Kogut, \href{http://arxiv.org/abs/arXiv:astro-ph/9607100}{{\tt
  arXiv:astro-ph/9607100}}.

\bibitem{Seager:1999bc}
S.~Seager, D.~D. Sasselov, and D.~Scott,
\href{http://arxiv.org/abs/astro-ph/9909275}{{\tt arXiv:astro-ph/9909275}}.

\bibitem{Jaeckel:2008fi}
J.~Jaeckel, J.~Redondo, and A.~Ringwald, {\em Phys. Rev. Lett.} {\bf 101}
  (2008)  131801,
\href{http://arxiv.org/abs/0804.4157}{{\tt arXiv:0804.4157 [astro-ph]}}.

\bibitem{Mirizzi:2009iz}
A.~Mirizzi, J.~Redondo, and G.~Sigl,
  \href{http://dx.doi.org/10.1088/1475-7516/2009/03/026}{{\em JCAP} {\bf 0903}
  (2009)  026},
\href{http://arxiv.org/abs/0901.0014}{{\tt arXiv:0901.0014 [hep-ph]}}.

\bibitem{Mirizzi:2009nq}
A.~Mirizzi, J.~Redondo, and G.~Sigl,
  \href{http://dx.doi.org/10.1088/1475-7516/2009/08/001}{{\em JCAP} {\bf 0908}
  (2009)  001},
\href{http://arxiv.org/abs/0905.4865}{{\tt arXiv:0905.4865 [hep-ph]}}.

\bibitem{Bartlett:1988yy}
D.~F. Bartlett and S.~Loegl,
{\em Phys. Rev. Lett.} {\bf 61} (1988)  2285--2287.

\end{thebibliography}

\providecommand{\href}[2]{#2}\begingroup\raggedright\endgroup

\end{footnotesize}


\end{document}